\documentclass[showpacs,aps,prl,twocolumn]{revtex4}

\pdfoutput=1
\usepackage{graphicx}
\usepackage{amssymb}
\usepackage{amsfonts,amsmath}

\begin{document}

\title{The geometrical nature of optical resonances in nanoparticles}

\author{F. Papoff$^\ast$ and B. Hourahine$^\ast$} \affiliation{SUPA,
  Department of Physics, University of Strathclyde, 107 Rottenrow,
  Glasgow G4 0NG, UK.(email: papoff@phys.strath.ac.uk,
  benjamin.hourahine@strath.ac.uk)}

\begin{abstract}
  We give a geometrical theory of resonances in Maxwell's equations
  that generalizes Mie formulae for spheres to any dielectric or
  metallic particle without sharp edges.  We show that the
  electromagnetic response of a particle is given by a set of modes of
  internal and scattered fields and reveal a strong analogy between
  resonances in nanoparticles and excess noise in unstable macroscopic
  cavities. We give examples of two types of optical resonances: those
  in which a single pair of internal and scattered modes become
  strongly aligned in the sense defined in this paper, and those
  resulting from constructive interference of many pairs of weakly
  aligned modes, an effect relevant for sensing. We demonstrate that
  modes can be either bright or dark depending on the incident field
  and give examples of how the excitation can be optimized.  Finally,
  we apply this theory to gold particles with shapes often used in
  experiments.
\end{abstract}


\maketitle

The interaction of light with wavelength sized particles has been
intensely investigated for more than a century, continuing to provide
interesting and surprising results. This coupling is essential in
single molecule spectroscopy and photon
processes~\cite{graham08a,aoki06a}, while interference between
different scattering channels of particles produce classical analogues
of quantum processes~\cite{lukyanchuk10a,liu09a}, and carefully
designed particles underpin the physical realization of
metamaterials~\cite{schuller10a,zhao09,pendry06a}. The basis all of
these effects is the strong variation of particle-light interaction
with wavelength.  For particles much larger than the wavelength of
light, resonances are described by closed orbits of light
rays~\cite{roll00a} inside the particle. This geometric approach
becomes less and less effective as the size of the particle decreases,
eventually requiring the solution of Maxwell's equations.  Mie-type
theories\cite{Mie1908,Han01}, based on symmetry and coordinate
separability, provide analytical description of resonances for a few
specific shapes of particle. For spheres, a resonance occurs when the
coefficient of one of the electric or magnetic multipoles in the field
expansion becomes infinite for particular values of the particle
radius, permittivity and susceptibility. In this framework, resonances
are not connected to anything similar to energy levels (eigenvalues of
the familiar hermitian operators of quantum mechanics), and do not
show an obvious geometrical interpretation.  Here we introduce an
hermetian operator to reveal the geometrical nature of resonances in
the Maxwell equations.

We consider metallic and dielectric particles without sharp edges and
of dimensions of the order of the wavelength of the incident field or
smaller, so that the interaction between light and matter inside the
particle is described by local macroscopic permittivity and
susceptibility. The tangential components of electric and magnetic
fields $E,H$ are continuous on passing through particle boundaries,
and the energy scattered by a particle flows towards infinity. We
use~\cite{holms09a} six component vectors $F = [E,H]^T$ for
electromagnetic fields; their projections, $f$, onto the surface of
the particle are surface fields that form a space $\mathcal{H}$ where
scalar products are defined in terms of surface integrals. In this
formalism the boundary conditions become
\begin{equation}  
  f^0=f^i-f^s,
\end{equation}
which has a simple geometrical meaning in $\mathcal{H}$: the
projection $f^0$ of the incident field onto the surface is equal to
the difference between the projections of the internal and scattered
fields, $f^i$ and $f^s$. This suggests that an external field with
small tangent components can excite large internal and scattered
surface fields provided that these cancel each other. This happens
when the ``angle'' between these two fields, and therefore their
difference, is small.

Importantly, for the particles considered here, sets of electric and
magnetic multipoles that are linearly independent and complete in
$\mathcal{H}$ can always be found~\cite{aydin86a,doicu06a}. Using
these sets we then form separate orthonormal bases $\{i_n\}$ and
$\{s_n\}$ for the projections of internal and scattered fields.  These
fields, which we call principal internal and scattering modes, are
orthogonal to all but at most one function in the other space.  For
spherical particles, each pair of modes corresponds to a pair of
electric or magnetic multipoles of Mie theory. For non-spherical
particles, principal modes are instead combinations of different
multipoles (although in some cases there can be dominant contributions
from a specific multipole). We can find the internal and scattered
fields for a particular incident field by decomposing this field into
a sum of pairs of modes $f^0 = \sum_n a^i_n {i}_n - a^s_n {s}_n $,
where $a^{i/s}$ are amplitudes of the internal and scattered principal
modes for that specific incident field.  The angle, $\xi_n$, between
$s_n$ and $i_n$ is defined as
\begin{equation}
  i_n \cdot s_n = \cos{(\xi_n)}, \label{pr_cos}
\end{equation}
where we choose the arbitrary phase factors so that the scalar product
is either positive or null. The terms on the right-hand side of
eq.~(\ref{pr_cos}) are the principal cosines introduced for the first
time by Jordan in 1875~\cite{jordan875}.  $\sin{(\xi_n)}$ is the
orthogonal distance between $s_n$ and $i_n$, and $\cos{(\xi_n)}$ their
statistical correlation~\cite{hannan61a}. These angles are invariant
under rotation and/or multiplication by a phase
factor~\cite{knyazev10a} and characterize the geometry of
the subspaces of the internal and scattered solutions in
$\mathcal{H}$.  This geometry is induced by the particular particle
through the surface integrals of the scalar products; its relevance to
scattering and resonances has not been previously realized. The
importance of the principal cosines is twofold: Theoretically they
provide analytic equations for the coefficients of the internal and
scattered principal modes, generalizing the Mie formulae and
clarifying the nature of all scattering channels of a particle.
Numerically, they allow us to reduce large matrices to their
sub-blocks and eliminate the need of numerical inversion for the
determination of the mode coefficients.

Minimization of $|f^0 - \sum_n a^s_n {s}_n + a^i_n {i}_n|$ leads to
the equation
\begin{equation}
\left[ \begin{array}{cc}
1 & C\\
C^{\dagger} & 1
\end{array} \right]
\left[ \begin{array}{c} {\bf a}^i \\ -{\bf a}^s \end{array} \right]=
\left[ \begin{array}{c} \mathcal{I}^{\dagger} f^0 \\ 
\mathcal{S}^{\dagger} f^0 \end{array}
  \right], \label{I_S_mat}
\end{equation}
where $\mathcal{I},\mathcal{S}$ are matrices whose columns are formed
by the principal modes $\{i_n\}$ and $\{s_n\}$, respectively, $1$ is the
identity and
${\bf a}^i $, ${\bf a}^s $ are the coefficients of the principal modes
in the field's expansion. The most important part of our theory is
that principal modes are coupled pairwise, i.e. that the matrix $C$ is
diagonal.  The interaction of particles with light can then be
interpreted in terms of eigenvalues and orthogonal eigenvectors
$w^{\pm}_n = (i_n \pm s_n)/\sqrt{2}$ of the hermitian operator in
Eq.(\ref{I_S_mat}).  However, away from the surface one measures
either internal or scattered fields, so we transform the
eigenfunctions, $\{w^{\pm}_n\}$, to find the coefficients of the
principal modes:
\begin{align}
  a^i_n &= \frac{i_n - \cos{(\xi_n)} s_n }{\sin^2{(\xi_n)}}\cdot f^0,
  \label{gen_mie_i} \\
  a^s_n &= - \frac{s_n - \cos{(\xi_n)} i_n }{\sin^2{(\xi_n)}}\cdot
  f^0.
  \label{gen_mie_s}
\end{align} 
Here $i^\prime_n=i_n - \cos{(\xi_n)} s_n$, $s^\prime_n=s_n -
\cos{(\xi_n)} i_n$ are bi-orthogonal to $i_n,s_n$ ($i^\prime_n \cdot
s_n =s^\prime_n \cdot i_n =0$) with $i^\prime_n \cdot i_n = s^\prime_n
\cdot s_n = \sin^2{(\xi_n)}$. $a^i_n,a^s_n$ in
Eqs.~(\ref{gen_mie_i},\;\ref{gen_mie_s}) are found by projecting the
incident field $f^0$ onto non-orthogonal vectors $i_n, s_n$ and
$\sin{(\xi_n)}$ is defined as the Peterman factor~\cite{new95a} that
gives the order of magnitude of transient gain and excess noise in
unstable cavity modes.  Therefore the presence of strongly aligned vectors
with $\sin{(\xi_n)}<<1$ is at the origin of large surface fields in
nanoparticles as well as large transient gain and excess
noise~\cite{firth05a} in macroscopic unstable cavities and dissipative
systems governed by non-hermitian operators~\cite{papoff08a}. However,
the physical origin of the non-orthogonality of $i_n, s_n$ is not
dissipation, but the fact that internal and scattered fields are not
modes of the hermitian operator and are correlated, even in the case
of non-absorbing particles.

These surface modes and their amplitudes fully specify the
electromagnetic behavior of the particle everywhere in space, for
example electric currents within the particle and its surface plasmons
can be found using the fields and the Ohm equation.
Eqs.~(\ref{gen_mie_i},\;\ref{gen_mie_s}) themselves have several
important consequences.  The left terms in the scalar products depend
exclusively on the particle, this allows us to disentangle its
properties from those due to the specific incident field. As an
example, we now consider resonances. The principal modes and scalar
products depend on the frequency dependent permittivity and
susceptibility of the particle, and therefore the principal angles
$\xi_n$ change with the frequency of incident light. Internal and
scattered coefficients diverge when the denominators of
eqs.~(\ref{gen_mie_i},\;\ref{gen_mie_s}) vanish. This happens when a
pair of normalized internal and scattering modes are parallel. For a
sphere the angular dependence of internal and scattered modes can be
factored out and the condition $i_n = s_n $ can be recast in terms of
the amplitude of the electric and magnetic components giving the Mie
resonance condition. For physical particles, the linear independence
and completeness of the principal modes makes perfect alignment
impossible. So, as with spherical particles~\cite{tribelsky06a},
actual resonances correspond to minimum angles ($\xi_n \ne 0$) of
pairs in $\mathcal{H}$. One important difference from spherical
particles exhibited by non-spherical structures is that the total flux
of energy scattered or absorbed (integrals of the Pointing vectors
over all directions) is given by the sum of principal mode
contributions plus interference terms between modes, which are absent
for spherical particles. Hence for non-spherical particles, strong
peaks in the efficiencies can be caused by constructive interference
within a group of modes. Furthermore,
Eqs.~(\ref{gen_mie_i},\;\ref{gen_mie_s}) show that modes can be
``dark'' for {\em specific} incident fields.  The largest internal
surface field is produced by an incident surface field parallel to
$i^\prime_n$: For well aligned modes, the coefficient $|a^i|,|a^s|$
are of the same order while, for weak resonances, $|a^s|$ is much
smaller than $|a^i|$, leading qualitatively different absorption and
differential scattering cross sections (DSCS). Similar effects happen
for an incident surface field parallel to $s^\prime_n$, exchanging
$|a^i|$ and $|a^s|$.

Regarding numerical calculations, we remark that $\sum_{n=1}^N a^i_n
I_n, \sum_{n=1}^N a^s_n S_n$ converge to the exact fields for $N
\rightarrow \infty$ at any point inside and outside the 
particle~\cite{rother02a,holms09a}, and $|f^0 +
\sum_{n=1}^N a^s_n s_n - a^i_n i_n|$ allows us to verify the accuracy
of the calculations for finite $N$~\cite{doicu06a}. Most importantly,
the form of Eqs.(\ref{gen_mie_i},\ref{gen_mie_s}) remains unchanged as
$N \rightarrow \infty$, even if $\theta_n,i_n,s_n$
change~\cite{knyazev10a}.

\begin{figure}
  \caption{\label{fig:discModes}(Color online) Calculated modes of the gold
    nanodisc. a) Scattering, absorption and extinction efficiencies,
    showing the strong resonance at 613~nm. b) The DSCS as a function
    of wavelength, illustrating the dipole nature of this particular
    resonance. c) and d) Excitation paths of scattered and internal
    fields over the principal angle landscape. e) and f) Near field
    plots of the dipole resonance disc viewed parallel and
    perpendicular to the electric field of the axial light.}
  \begin{center}
    \includegraphics[width=.49\textwidth,clip=true]{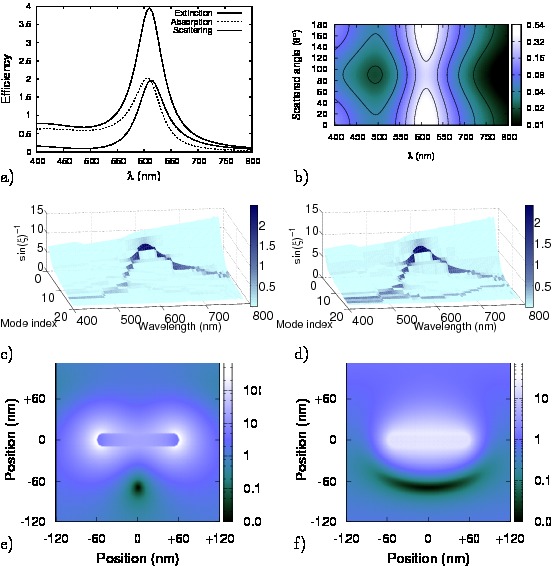}
  \end{center}
\end{figure}

We now consider light interaction with some specific gold particles,
using a fitted dielectric
function~\cite{Etchegoin06}. Fig.~\ref{fig:discModes} shows calculated
near and far field properties of a rounded nanodisc (height 20~nm and
diameter 120~nm), illuminated axially by plane-wave light. We find a
strong dipole resonance (fig.~\ref{fig:discModes}a,b) at 613~nm, in
agreement with a previous experimental and numerical
investigation~\cite{aizpurua03a}.  Figs.~\ref{fig:discModes}c,d) show
the landscape of principal mode angles, which are intrinsic properties
of the particle, as a function of mode index and wavelength of
incident light.  The ``height'' of this landscape is $\sin({\xi})^{-1}$,
the largest value of eq.~\ref{gen_mie_i} and eq.~\ref{gen_mie_s} for  
$|f^0|=1$, while the amplitudes of the internal and scattered
principal modes due to the {\em specific} axial incident field are
shown overlaid on top. These {\em excitation paths} demonstrate that
the observed resonance is a Mie-like single principal cosine pair,
which comes into maximum alignment (its smallest value of $\xi$) at
the resonance. Other strongly aligned dark modes are also present, but
not excited by this particular incident field. The internal field also
contains a second more weakly aligned excited mode, where its counterpart 
in the scattering field is not
excited. This mode and of the resonant mode
have quasi-degenerate principal angles on the shorter wavelength side
of its peak; this strengthens the internal field and explains the
appreciable asymmetry of the absorption efficiency.  The near field
around the particle at resonance (fig.~\ref{fig:discModes}e,f) shows
that the resonance is an electric dipole. For non axial incidence, the
main peak is smaller due to a weaker coupling with the resonant mode.

\begin{figure}
  \caption{\label{Fig:rodModes0}(Color online) Calculated optical modes of the
    rounded nanorod when illuminated axially. a) Scattering,
    absorption and extinction efficiencies showing the 205~nm mode and
    weaker absorption peak at 486~nm. b) DSCS showing the strong
    forward scattering of light by the particle. c) and d) Scattered and
    internal excitation paths, showing the composite nature of the
    modes.}
  \begin{center}
    \includegraphics[width=.49\textwidth,clip=true]{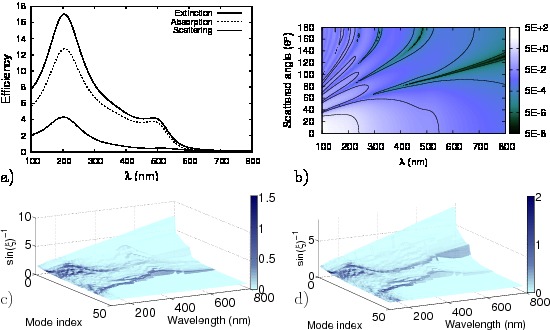}
  \end{center}
\end{figure}
Fig.~\ref{Fig:rodModes0}a,b) shows the calculated optical efficiencies
and DSCS of a 480~nm long rod with diameter 40~nm, illuminated axially
by plane-wave light.  There is a strong resonance at 205~nm and a
weaker absorption peak at 486~nm (an extremely weak corresponding
scattering feature is present at $\sim$515~nm). The DSCS demonstrates
that this particle strongly scatters this incident light forward,
particularly at short wavelengths.  The peak at 205~nm is not a single
mode, but instead the excitation amplitudes of the principal modes
(fig~\ref{Fig:rodModes0}c,d) show it to be due to constructive
interference of contributions from a group of several weakly aligned
principal mode pairs. Collectively these modes very efficiently
extract energy from the incident field. Such multimode resonances are
potentially useful for sensing applications because interference
between different scattering channels leads to enhanced sensitivity to
perturbations near to the scattering surfaces.  The absorption feature
at 486~nm is also due to a group of mode pairs, in which only
the internal modes are excited, all becoming more strongly aligned at 
around this
resonance.
\begin{figure}
  \caption{\label{Fig:rodModes90}(Color online) The particle from
    fig.~\ref{Fig:rodModes0}, but illuminated from the size with an
    incident light polarization of 45$^\circ$ with respect to its long
    axis. a) Far field scattering efficiencies showing the presence of
    both a broad feature similar to Fig.~\ref{Fig:rodModes0} at around
    200--450~nm and also a sharp resonance at 676~nm. b) The DSCS for
    equatorial illumination c) and d) Scattered and internal
    excitation paths; the principal angle landscape includes
    additional modes that cannot be excited by symmetry for axial
    incidence (cf.\ Fig.~\ref{Fig:rodModes0}). The sharp mode is again
    a single well aligned principal pair, the best aligned of three
    (and the only one excited by this field). e) and f) Near field
    for the 480~nm rod, shown at the broad feature (207~nm) and at the
    ``waveguide'' Mie-like mode at 676~nm.}
  \begin{center}
    \includegraphics[width=.49\textwidth,clip=true]{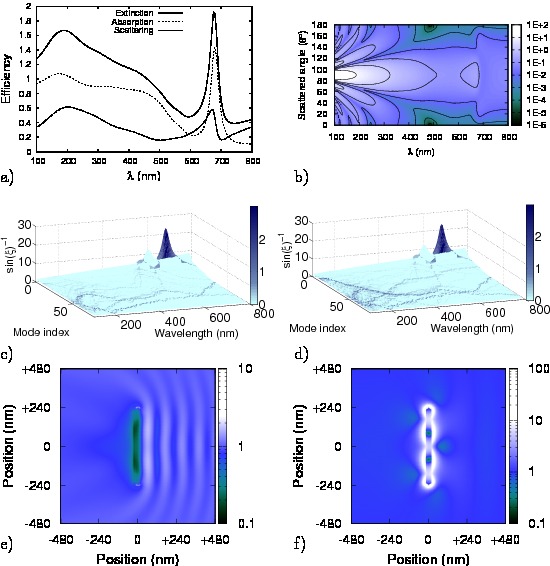}
  \end{center}
\end{figure}
Fig.~\ref{Fig:rodModes90} shows the same particle illuminated
equatorially with an incident light polarization of 45$^\circ$ with
respect to its long axis.  Analyzing other particles with the same
diameter but varying length, we find that the broad feature at around
200--450~nm, containing structures similar to the composite modes of
Fig.~\ref{Fig:rodModes0}, is insensitive to the particle length. It
also shows no clear hot or cold spots (see fig~\ref{Fig:rodModes90}e).
The sharp resonance at 676~nm shifts with the length of the rod and
its surface field has the strong nodal local structure
(Fig.~\ref{Fig:rodModes90}f) of a ``waveguide'' mode on the long
axis. The excitation paths (Fig.~\ref{Fig:rodModes90}c,d) are shown
including additional modes that cannot be excited by symmetry for
axial incidence (cf.\ Fig.~\ref{Fig:rodModes0}). As with the disc, the
sharp resonance is again a single well aligned principal pair, the
best aligned (and the only one excited by this field) of three such
waveguide modes that shift with the length of the particle.


In summary, we provide a geometrical explanation for resonances that generalizes
Mie formulas for spheres to all isotropic  particles without sharp edges and
reveal strong analogies between resonances in nanoparticles and excess noise 
in macroscopic cavities.  We show that there are sharp resonances
caused by strong alignment between one internal and one scattered modes and
broad resonances due to several pairs of internal and scattered modes with weak,
but similar, alignment.  Our equations indicate that any observation depends
both on the structure of the principal modes, which are an intrinsic property of
the particle, and the way an incident field couples to these modes.  In
particular, there can be resonant modes that are essentially dark with respect
to a given incident field. The amplitudes of the principal modes specify the
field both inside and outside the particle.  This geometric approach is
applicable to a large class of processes, where the interaction between the
system and the environment is described by a set of functions complete at the
boundary of the system, including other scattering processes such as of acoustic
or electron waves, and coupling to optical cavities.


\end{document}